\documentclass[review]{elsarticle}

\journal{Journal of \LaTeX\ Templates}

\biboptions{authoryear}

\usepackage{bm}

\begin{document}

\begin{frontmatter}

\title{Possible relation between pulsar rotation and evolution of magnetic inclination}

\author{Tian, Jun}
\address{Department of Physics, The University of Hong Kong, Pokfulam Road, Hong Kong}


\ead{juntian@connect.hku.hk}


\begin{abstract}
The pulsar timing is observed to be different from predicted by a simple magnetic dipole radiation. We choose eight pulsars whose braking index was reliably determined. Assuming the smaller values of braking index are dominated by the secular evolution of the magnetic inclination, we calculate the increasing rate of the magnetic inclination for each pulsar. We find a possible relation between the rotation frequency of each pulsar and the inferred evolution of the magnetic inclination. Due to the model-dependent fit of the magnetic inclination and other effects, more observational indicators for the change rate of magnetic inclination are needed to test the relation.
\end{abstract}


\end{frontmatter}


\section{Introduction}

Pulsars are highly magnetized rotating neutron stars with beams of emission that are observed as pulses. The slowdown of rotation is manifest in the actual arrival times of pulses. Usually the loss of kinetic rotational energy is understood to be transformed to electromagnetic radiation. 
In the idealized model, neutron stars slow down because of the generation of magnetic dipole radiation.
In this case, the slowdown of frequency is given by

\begin{equation}
\dot{\nu}=-\frac{8\pi^{2}}{3c^{3}}\frac{M^{2}sin^{2}\alpha}{I}\nu^{3}
\end{equation}

where $M$ is the magnetic dipole moment, $I$ the moment of inertia, $\alpha$ the inclination angle between the magnetic and rotation axis. In the canonical model $M$, $I$ and $\alpha$ are constants, and a dependency $\dot{\nu}\propto\nu^{3}$ for the slowdown of pulsars would be expected. 

However the observed dependence of $\dot{\nu}$ on $\nu$ differs from predicted by the magnetic dipolar radiation. The power-law slowdown can be written in a more general form

\begin{equation}
\dot{\nu}\propto\nu^{n}
\end{equation}

where $n$ is known as the braking index. Reliable values of $n$ were obtained for a few pulsars (see Table 1). Most of the braking index values are less than 3 except for PSR J1640-4631 which was recently measured to be 3.15 \citep{Arc16}.

The discrepancy of observed braking index challenge the canonical model, and different approaches have been tried to improve it\citep{BR88, AH97, Mel97, CS06, MMF12, KT15}. One way is to consider the time evolution of the three constants in equation (1) $M$, $I$ and $\alpha$. The evolution of magnetic inclination is most intriguing since the report of an increasing $\alpha$ in Crab pulsar though very slow\citep{AF13}. Most excitingly, the observed braking index value of 2.51\citep{Ly93} can be remarkably explained if the departure from classic slowdown is solely attributed to a secular change in $\alpha$. Inspired by this success, we tried to interpret the small values of braking index for the eight pulsars in Table 1 being dominated by a secular anti-alignment between magnetic and rotation axis. From equation (1), $\dot{\nu}\propto\nu^{3}sin^{2}\alpha$. So the modified braking index is given by 

\begin{equation}
n=3+2\nu/\dot{\nu}\times\dot{\alpha}/tan\alpha
\end{equation}

If the magnetic inclination angle is well constrained, we can derive its change rate $\dot{\alpha}$ from this equation. One method to constrain the viewing geometry is to fit the observed polarization position angle swing by the Rotating Vector Model\citep{RC69}. However for the pulsars in Table 1 no reliable parameters of viewing geometry are available due to extreme uncertainties. So the $\alpha$ values adopted in this work depend on different models. Most of them were derived from the fit of high-energy light curves.

\section{Review of model dependent $\alpha$}

For the Crab pulsar the shape of the beam has been modeled by several authors\citep{DR03, Har08, Wat09, DQW12}, and the estimated range of $\alpha$ is between $45^{\circ}$ and $70^{\circ}$. A model of high-energy light curves whose double peaks are supposed to arise from a crossing two caustics and associated with different magnetic poles produces generic features consistent with observed pulsar light curves, and particularly with $\alpha=60^{\circ}$ the model is well suited for the light curve of the Crab pulsar and $\alpha=70^{\circ}$ for the Vela pulsar\citep{DR03}. A 3D model of emission from optical to gamma-ray band originating from the high-altitude slot gap accelerator can reasonably well reproduce the Crab pulsar profile and spectrum with $\alpha=45^{\circ}$\citep{Har08}. A simulation of the beaming pattern and light curves for young spin-powered pulsars leads to an estimate of $\alpha$ large as $70^{\circ}$ with an outer gap geometry and a range between $55^{\circ}$ and $60^{\circ}$ with two pole caustic\citep{Wat09}. In the same simulation $\alpha$ for the Vela pulsar is estimated to be $75^{\circ}$ configured with outer gap and between $62^{\circ}$ and $68^{\circ}$ with two pole caustic. A fit of the phase-averaged spectrum and phase-resolved spectra of the Crab pulsar, which was best modeled by the annular gap emission shows an inclination angle of $45^{\circ}$\citep{DQW12}.

For PSR B0540-69, based on the 3D outer magnetosphere model of pulsars\citep{CRZ00} the calculated light curve and spectrum assuming a magnetic inclination of $50^{\circ}$ are similar to the observed data\citep{ZC00}. Using the same model, a magnetic inclination of $60^{\circ}$ is consistent with the observed data for PSR B1509-58. In a later simulation\citep{TC07} a smaller inclination angle $\alpha=30^{\circ}$ for PSR B0540-69 was adopted to reproduce a pulse profile more consistent with observation.

For the Vela pulsar a fit of the phase-averaged spectrum whose gamma-ray emission was modeled by photon-photon pair process in the outer gap gives $\alpha=71\pm1.1^{\circ}$\citep{LJZ13}. In the same case $\alpha=70^{\circ}$ was estimated for PSR J1833-1034 by calculating the best exponential cutoff power-law fit since the observed spectral data were not available. It can be seen the observed spectrum for this pulsar cannot be reproduced well in this model. Another model of the retarded vacuum dipole field in conjunction with standard outer gap emission geometry was applied to fit the gamma-ray light curve of the Vela pulsar and present an optimal solution $\alpha=78\pm1^{\circ}$\citep{BVH16}.

The polarization profiles are available for PSR J1119-6127 and PSR B1509-58\citep{RWJ15}. The fit of the position angles to the Rotating Vector Model constrains the viewing geometry. For PSR J1119-6127 two favored solutions $\alpha=9.1^{\circ}$ or $6.9^{\circ}$ were given corresponding to the rotating radio transients like components outside and inside the open field line region respectively(values of $\alpha>90^{\circ}$ have been mapped into $0<\alpha<90^{\circ}$). For PSR B1509-58 the favored solution was $\alpha=13.7^{\circ}$. After correcting the values of $\alpha$ based on an intrinsic sinusoidal distribution, the favored solution for PSR J1119-6127 was $\alpha=21^{\circ}$ or $16^{\circ}$ and for PSR B1509-58 $\alpha=30^{\circ}$\citep{RWJ15_2}. Another fit of the observed gamma-ray spectrum and energy dependent light curves for PSR B1509-58 under the outer gap model found $\alpha=20^{\circ}$\citep{WTC13}.

For the GeV-quite soft gamma-ray PSR J1846-0258, the soft spectrum and single-peak light curve could be well explained by a model suggesting the emissions produced via synchrotron radiation of the electron-positron pairs which are created by the inward gamma rays interacting with the strong magnetic field near the polar cap region with the viewing geometry of $\alpha=10^{\circ}$\citep{WNT14}. For high-magnetic radio pulsar PSR J1734-3333, the calculations of magnetic inclination suggest $\alpha=6^{\circ}$ if the line of sight passes through the center of the emission cone and $\alpha=21^{\circ}$ if polarization data is taken into account\citep{NM16}. In the same calculation, we can see for PSR J1119-6127 $\alpha\sim6^{\circ}$ and $17^{\circ}$ via the two methods respectively, and for PSR B1509-58 $\alpha\sim3^{\circ}$ and $10^{\circ}$.

PSR J1640-4631 is the only pulsar with a reliable braing index larger than three. The braking index $3.15\pm0.03$\citep{Arc16} cannot be attributed to the anti-alignment of magnetic axis. It can be explained by the plasma-filled magnetosphere model\citep{Spi06, Phi14} for two different inclination angles, $18.5^{\circ}\pm3^{\circ}$ and $56^{\circ}\pm4^{\circ}$\citep{Eks16}. The smaller value is favored by the single-peak pulse profile. The rate of decrease of the inclination angle was found to be $(-0.23\pm0.05)^{\circ}century^{-1}$.

\section{Possible relation between $\dot{\alpha}$ and $\nu$}

For each of the nine pulsars differently favored values of the magnetic inclination angle have been reported by several authors as listed in Table 2. 
The orientation of magnetic axis for different pulsars range from nearly alignment with the rotation axis to orthogonal. The favored values of $\alpha$ for one pulsar made by different methods can be very large.

The discrepancy of braking index from three for the nine pulsars is shown in Figure 1.
There is no significant correlation with the spin frequency as shown in the figure. This is suggested by the large probable error in the correlation coefficient of $-0.1\pm0.2$. Note PSR J1640-4631 is ignored in the analysis of correlation.

Under the assumption that the discrepancy of braking index of timing from the canonical model is dominated by the secular evolution of the magnetic inclination, the inferred values of $\dot{\alpha}$ were obtained as listed in Table 2 except for PSR J1640-4631 whose $\dot{\alpha}$ was found by \citet{Eks16}. Many methods have been invented to constrain the magnetic inclination. For PSR J1119-6127 and PSR B1509-58 there are as many as six favored solutions while only one solution is available for PSR J1833-1034 and PSR J1846-0258. All the possible values of $\dot{\alpha}$ are plotted in Figure 2 against the spin frequency of each pulsar. There are large scatterings of $\dot{\alpha}$ for the pulsars whose values of $\alpha$ were measured more than once.

The smallest values of $\dot{\alpha}$ concentrate on the low-frequency end of Figure 2 while largest values on the high-frequency end. There seems to be an overall trend of $\dot{\alpha}$ increasing with pulsars' rotation speed. For the pulsars with more than one value of $\dot{\alpha}$, we divided the values into three cases: maximum, minimum and average. The correlation coefficients between $\dot{\alpha}$ and $\nu$ for the three cases are respectively $0.94\pm0.03$, $0.73\pm0.11$ and $0.89\pm0.05$. We tried to fit the possible positive relation between $\dot{\alpha}$ and $\nu$ with a straight line. As shown in Figure 2 the two dashed line correspond to the fitting curve of the maximum and minimum values of $\dot{\alpha}$, and the solid line is for the average values of $\dot{\alpha}$. The two dotted lines represent the bounds for the fit of the averaged $\dot{\alpha}$. The error of the fitted coefficients of the linear dependence are determined by the variance of the fitting residuals within $95\%$ confidence interval. The two dotted lines correspond to the extreme estimates of $\dot{\alpha}$.
We can see for the Crab pulsar the increasing rate of the magnetic inclination inferred from the slowly increasing component separation in the radio pulse profile\citep{AF13} lies well within the bounds.

\section{Discussion}



The relation between $\dot{\alpha}$ and $\nu$ are easily affected by other parameters. As is well known, there is a strong correlation between the frequency and its derivative for all known pulsars. To see whether this relation exists among the small population of the eight pulsars and affects the deduced relation in Figure 2, we plotted the two parameters in Figure 3. There seems to be a linear dependence of the two parameters which is suggested by the correlation coefficient of $-0.87\pm0.06$. This may be induced by the similar characteristic age of the young pulsars considered here. Therefore the relation in Figure 2 is not necessarily true and needs testing with pulsars of various characteristic ages.

The possible relation between the evolution of the magnetic inclination and the rotation frequency presented here is based on the assumption that the various timing behavior unexpected by the magnetic dipole radiation is due to a secular change of magnetic inclination. Many other effects can also result in the same timing behavior such as the evolution of the magnetic field\citep{ZX12} and an interaction between fall-back disk and magnetic field\citep{CL16}. Taking into consideration these effects the actual evolution of the magnetic inclination might differ very much from the values calculated here.

The poor measurement of the magnetic inclination is another obstacle to a simple relation. Different models present a large scattering of magnetic inclination angles. The direct determination is only available for the Crab pulsar. We would expect more Crab-like cases to test the possible relation between the change rate of magnetic inclination and rotation frequency.

The slow evolution of magnetic inclination like in the Crab pulsar is difficult to detect. If it's positively related to the rotation speed as shown in Figure 2 among a large population of pulsars, we would expect a faster transfer of the magnetic pole for some fastest rotating pulsars, which may be easier to observe.

\begin{table}[h]
\centering
\caption{Rotation frequencies for pulsars with known braking indices. Time derivatives are denoted by a dot.}
\begin{tabular}{cccccc}
\hline
PSR & n & $\nu(s^{-1})$ & $\dot{\nu}(\times10^{-11}s^{-2})$ & reference \\
B0531+21(Crab) & 2.51(1) & 30.2254371 & -38.6228 & \citet{Ly93} \\
B0540-69 & 2.14(1) & 19.8344965 & -18.8384 & \citet{Li07} \\
B0833-45(Vela) & 1.4(2) & 11.2 & -1.57 & \citet{Ly96} \\
J1119-6127 & 2.684(2) & 2.4512027814 & -2.415507 & \citet{WJE11} \\
B1509-58 & 2.839(1) & 6.633598804 & -6.75801754 & \citet{Li07} \\
J1734-3333 & 0.9(2) & 0.855182761 & -0.166702 & \citet{Es11} \\
J1833-1034 & 1.857(1) & 16.15935712 & -5.275113001 & \citet{RGL12} \\
J1846-0258 & 2.65(1) & 3.0782148166 & -6.71562 & \citet{Li07} \\
J1640-4631 & 3.15(3) & 4.84341082106 & -2.280775 & \citet{Arc16} \\
\hline
\end{tabular}
\end{table}

\begin{table}[h]
\centering
\caption{The different values of magnetic inclination depend on different models. The corresponding change rate of magnetic inclination is supposed to produce the braking index smaller than three. Note for PSR J1640-4631 the value of $\dot{\alpha}$ is predicted by the plasma-filled magnetosphere model in the reference.}
\begin{tabular}{ccccc}
\hline
PSR & $\alpha(degree)$ & $\dot{\alpha}(10^{-12}rad/s)$ & reference \\
B0531+21(Crab) & 60 & 5.35 & \citet{DR03} \\
 & 45 & 3.089 & \citet{Har08, DQW12} \\
 & 70 & 8.486 & \citet{Wat09} \\
 & 55-60 & 4.411-5.35 & \citet{Wat09} \\
 B0540-69 & 50 & 4.853 & \citet{ZC00} \\
 & 30 & 2.351 & \citet{TC07} \\
 B0833-45(Vela) & 70 & 3.076 & \citet{DR03} \\
 & 75 & 4.178 & \citet{Wat09} \\
 & 62-68 & 2.106-2.771 & \citet{Wat09} \\
 & 69.9-72.1 & 3.059-3.466 & \citet{LJZ13} \\
 & 77-79 & 4.849-5.76 & \citet{BVH16} \\
 J1119-6127 & 9.1 & 0.2494 & \citet{RWJ15} \\
 & 6.9 & 0.1884 & \citet{RWJ15} \\
 & 21 & 0.5977 & \citet{RWJ15_2} \\
 & 16 & 0.4465 & \citet{RWJ15_2} \\
 & 6 & 0.1636 & \citet{NM16} \\
 & 17 & 0.476 & \citet{NM16} \\
 B1509-58 & 60 & 1.412 & \citet{ZC00} \\
 & 13.7 & 0.199 & \citet{RWJ15} \\
 & 30 & 0.471 & \citet{RWJ15_2} \\
 & 20 & 0.297 & \citet{WTC13} \\
 & 3 & 0.043 & \citet{NM16} \\
 & 10 & 0.144 & \citet{NM16} \\
 J1734-3333 & 6 & 0.2151 & \citet{NM16} \\
 & 21 & 0.7857 & \citet{NM16} \\
 J1833-1034 & 70 & 5.1258 & \citet{LJZ13} \\
 J1846-0258 & 10 & 0.67157 & \citet{WNT14} \\
 J1640-4631 & $18.5\pm3$ & $-(1.3\pm0.3)$ & \citet{Eks16} \\
\hline
\end{tabular}
\end{table}


\begin{figure}
\begin{center}
\includegraphics[width=1.0\textwidth]{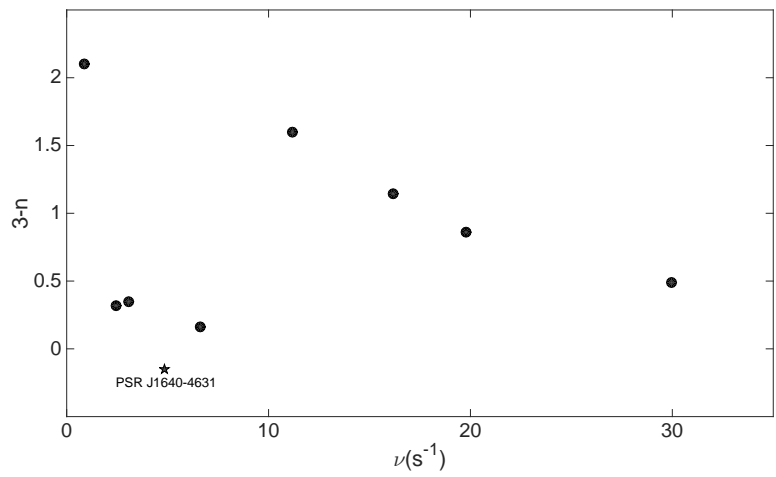}
\end{center}
\caption{\small The discrepancy of braking index from 3 for the eight pulsars are plotted as dotted points, while the only pulsar with larger than 3 braking index, PSR J1640-4631 is specially marked with a star in the bottom.}
\end{figure}

\begin{figure}
\begin{center}
\includegraphics[width=1.0\textwidth]{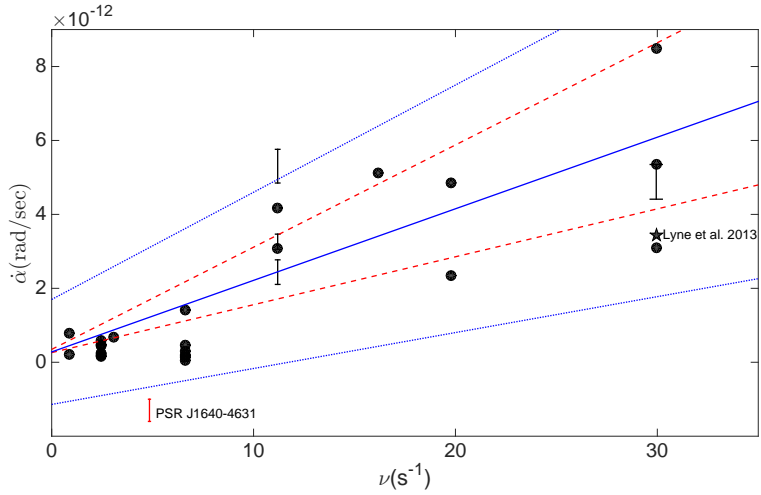}
\end{center}
\caption{\small The relation between the change rate of the magnetic inclination and spinning frequency. The dotted points at the same frequency correspond to the different values of the magnetic inclination determined by different authors. The two dashed lines are linear fits to the points with maximum and minimum known $\dot{\alpha}$ respectively. The solid line is the fit to the average values of $\dot{\alpha}$ bounded by the two dotted lines with $95\%$ confidence level. The star point represents the first observation of the secular change of the magnetic inclination of the Crab pulsar. The large braking index of PSR J1640-4631 is assumed to be induced by the alignment of magnetic and rotation axis, so its $\dot{\alpha}$ is negative. This outlier is specially marked with red color and not included in analyzing the above relation.}
\end{figure}

\begin{figure}
\begin{center}
\includegraphics[width=1.0\textwidth]{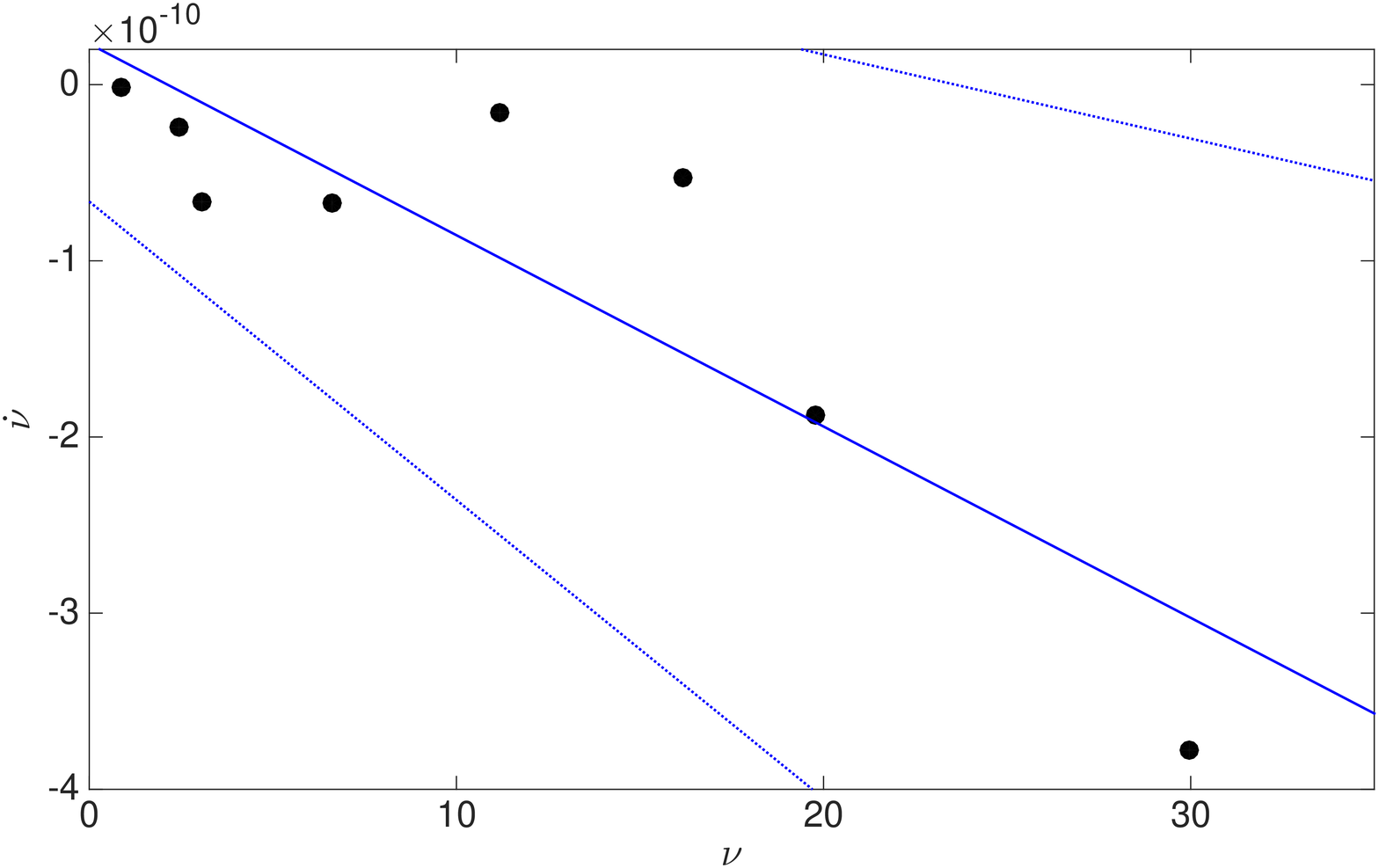}
\end{center}
\caption{\small The derivative of frequency plotted against the frequency for the eight pulsars. The solid line is a linear fit to the points with $95\%$ confidence bounded by two dotted lines.}
\end{figure}


\begin{thebibliography}{}

\bibitem[Archibald et al.(2016)]{Arc16} Archibald R. F. et al. 2016, ApJ, 819, L16

\bibitem[Blandford $\&$ Romani(1988)]{BR88} Blandford R. D., Romani R. W. 1988, MNRAS, 234, 57P 

\bibitem[Allen $\&$ Horvath(1997)]{AH97} Allen M. P., Horvath J. E. 1997, ApJ, 488, 409

\bibitem[Melatos(1997)]{Mel97}Melatos A. 1997 MNRAS, 288, 1049

\bibitem[Contopoulos $\&$ Spitkovsky(2006)]{CS06}Contopoulos I., Spitkovsky A. 2006, ApJ, 643, 1139

\bibitem[Magalhaes, Miranda $\&$ Frajuca(2012)]{MMF12}Magalhaes N. S., Miranda T. A., Frajuca C. 2012, ApJ, 755, 54

\bibitem[Kou $\&$ Tong(2015)]{KT15}Kou F. F., Tong H. 2015, MNRAS, 450, 1990

\bibitem[Lyne et al.(2013)]{AF13}Lyne A., Graham-Smith F., Weltevrede P., et al. 2013, Science, 342, 598

\bibitem[Lyne, Pritchard $\&$ Smith(1993)]{Ly93}Lyne A. G., Pritchard R. S., Smith F. G. 1993, MNRAS, 265, 1003

\bibitem[Radhakrishnan $\&$ Cooke(1969)]{RC69}Radhakrishnan V., Cooke D. J. 1969, Astrophys. Lett., 3, 225

\bibitem[Dyks $\&$ Rudak(2003)]{DR03}J. Dyks, B. Rudak 2003, ApJ, 598, 1201

\bibitem[Harding et al.(2008)]{Har08}A. K. Harding, J. V. Stern, J. Dyks, M. Frackowiak 2008, ApJ, 680, 1378

\bibitem[Watters et al.(2009)]{Wat09}K. P. Watters, R. W. Romani, P. Weltevrede, S. Johnston 2009, ApJ, 695, 1289

\bibitem[Du, Qiao $\&$ Wang(2012)]{DQW12}Y. J. Du, G. J. Qiao, W. Wang 2012, ApJ, 748, 84

\bibitem[Cheng, Ruderman $\&$ Zhang(2000)]{CRZ00}Cheng K. S., Ruderman M., Zhang L. 2000, ApJ, 537, 964

\bibitem[Zhang $\&$ Cheng(2000)]{ZC00}Zhang, L. $\&$ Cheng, K. S. 2000, A$\&$A, 363, 575

\bibitem[Takata $\&$ Chang(2007)]{TC07}J. Takata $\&$ H. K. Chang 2007, ApJ, 670, 677

\bibitem[Li, Jiang $\&$ Zhang(2013)]{LJZ13}X. Li, Z. J. Jiang $\&$ L. Zhang 2013, ApJ, 765, 124

\bibitem[Barnard, Venter $\&$ Harding(2016)]{BVH16}Barnard, M., Venter, C., Harding, A. K. 2016, ApJ, 832, 107

\bibitem[Rookyard, Weltevrede $\&$ Johnston(2015a)]{RWJ15}S. C. Rookyard, P. Weltevrede $\&$ S. Johnston 2015, MNRAS, 446, 3367

\bibitem[Rookyard, Weltevrede $\&$ Johnston(2015b)]{RWJ15_2}S. C. Rookyard, P. Weltevrede $\&$ S. Johnston 2015, MNRAS, 446, 3356

\bibitem[Wang, Takata $\&$ Cheng(2013)]{WTC13}Wang, Y., Takata, J., Cheng, K. S. 2013, ApJ, 764, 51

\bibitem[Wang et al.(2014)]{WNT14}Wang, Y., Ng, C. W., Takata, J. et al. 2014, MNRAS, 445, 604

\bibitem[Nikitina $\&$ Malov(2017)]{NM16}Nikitina, E. B. $\&$ Malov, I. F. 2017, Astronomy Report, 61, 591

\bibitem[Zhang $\&$ Xie(2012)]{ZX12}Shuang-Nan Zhang $\&$ Yi Xie 2012, ApJ, 757, 153

\bibitem[Chen $\&$ Li(2016)]{CL16}Chen, Wen-Cong $\&$ Li, Xiang-Dong 2016, MNRAS, 455, 87

\bibitem[Livingstone et al.(2007)]{Li07}Livingstone M. A., Kaspi V. M., Gavriil F. P. et al. 2007, Astrophys. Space Sci., 308, 317

\bibitem[Lyne et al.(1996)]{Ly96}Lyne A. G., Pritchard R. S., Graham-Smith F. 1996, Nature, 381, 497

\bibitem[Weltevrede, Johnston $\&$ Espinoza(2011)]{WJE11}Weltevrede P., Johnston S., Espinoza C. M. 2011, MNRAS, 411, 1917

\bibitem[Espinoza et al.(2011)]{Es11}Espinoza C. M., Lyne A. G., Kramer M. et al. 2011, ApJ, 741, 13

\bibitem[Roy et al.(2012)]{RGL12}Roy J., Gupta Y., Lewandowski W. 2012, MNRAS, 424, 2213

\bibitem[Spitkovsky(2006)]{Spi06}Spitkovsky, A. 2006, ApJL, 648, L51

\bibitem[Philippov et al.(2014)]{Phi14}Philippov, A., Tchekhovskoy, A. $\&$ Li, J.G. 2014, MNRAS, 441, 1879

\bibitem[Eksi et al.(2016)]{Eks16}Eksi, E.Y., Andac, A.C. et al. 2016, ApJ, 823, 34
\end{thebibliography}
\end{document}